\documentclass[aps,pra,preprint,groupedaddress,nofootinbib,
floatfix,amsmath,amssymb,showkeys]{revtex4-2}
\usepackage{graphicx}
\usepackage{multirow}

\begin{document}


\title{Wavefronts, light rays and caustic associated with the refraction of a plane wave by a conospherical lens}

\author{Jos\'e Israel Galindo-Rodr\'iguez}
\email[E-mail:]{israelgalrod@gmail.com}
\affiliation{Facultad de Ciencias F\'isico Matem\'aticas de la Benem\'erita Universidad Aut\'onoma de Puebla, Apartado Postal 1152, 72001, Puebla, Pue., M\'exico.}
\author{Gilberto Silva-Ortigoza}
\affiliation{Facultad de Ciencias F\'isico Matem\'aticas de la Benem\'erita Universidad Aut\'onoma de Puebla, Apartado Postal 1152, 72001, Puebla, Pue., M\'exico.}

\date{\today}

\begin{abstract}
The aim of the present work is to introduce a lens whose faces are a conical surface and a spherical surface. We illuminate this lens by a plane wavefront and its associated refracted wavefronts, light rays and caustic are computed. We find that the caustic has two branches. The first is constituted by two segments of a line, one part of this caustic is real and the other one virtual. The second branch of the caustic is a two-dimensional surface with a singularity of the cusp ridge type. It is important to remark that the two branches of the caustic are disconnected. Because of this property we believe that using this optical element one could generate a scalar optical accelerating beam in the region where the caustic is a two-dimensional surface of revolution, and at the same time a scalar optical beam with similar properties to the Bessel beam of zero order in the region were the real caustic is a segment of a line along the optical axis.
\end{abstract}

\keywords{Geometrical Optics, Lenses, Caustics.}

\maketitle

\section{Introduction}

In 1954 McLeod\,\cite{McLeod} introduced a new class of optical elements, these were called axicons. There are different kinds of axicons, but the most important one is a glass cone. At that time one of its possible applications was in a telescope\,\cite{McLeod2}. In this case, the usual spherical objective is replaced by a cone. This axicon telescope is in focus for targets from a foot or so to infinity without the necessity of moving any parts. It can be used to view simultaneously two or more small sources placed along the line of sight.

One of the most important properties of the axicon lens is its caustic, which is a focal line when it is illuminated by a plane wave evolving along the axis of symmetry of the lens. Because of this property, the axicon lens has been used for different purposes. In optical coherence tomography, axial and lateral resolutions are determined by the source coherence length and  the  numerical  aperture  of  the  sampling  lens,  respectively. Whereas  axial  resolution  can  be  improved by  use of  a  broadband  light  source,  there is  a trade-off  between  lateral  resolution and  focusing  depth  when conventional  optical  elements  are  used. Ding \textit{et al}.\,\cite{Zhihua Ding} incorporated an  axicon  lens  into  the  sample arm  of  an  interferometer  to  overcome  this  limitation. Burvall \textit{et al}.\,\cite{Anna Burvall} presented the design of a cemented double-lens axicon made from spherical surfaces only. This axicon lens is based on the deliberate use of the spherical aberration of the surfaces and has a narrow, extended line focus of relatively constant width. Golub\,\cite{Golub} had introduced the fraxicon, this element consists of concentric prism-like grooves with an apex angle equal to that of a bulk axicon and whose action relies  on  refraction. Cabrini \textit{et al}.\,\cite{Cabrini}  by using an axicon lens showed that it is possible to make efficient optical tweezers. B\'elanger and Rioux\,\cite{Pierre} combined an axicon and a lens to form an optical system producing a ring-shaped pattern to show that when a lens–axicon combination is illuminated by a Gaussian beam, the transverse distribution of the focal ring is also a Gaussian distribution. Tanaka and Yamamoto\,\cite{Tanaka} studied the aberrations produced by an axicon and a lens. They found that the beam spot of an axicon is affected only by astigmatism and not by coma, while the point created by a lens is blurred by both astigmatism and coma. This implies that in the scanning optical system that incorporates an axicon it is easy to correct the aberrations and achieves high-resolution and high contrast imaging with a wide field of view. Zheng \textit{et al}.\,\cite{Zheng} demonstrated experimentally an efficient electron axicon lens using a magnetic vortex. In particular, the axicon provides the most efficient method for realizing generally diffractionless beams\,\cite{McGloin}.

It is important to remark that there are few works on the axicon lens using geometrical optics\,\cite{McLeod, McLeod2, Rayces, Juarez-Reyes}. Rayces\,\cite{Rayces} studied, in a qualitative way, the formation of axicon images. To be more precise, he considered a simple refracting axicon illuminated by a point light source, placed on and off the optical axis, and described the form of the caustic associated with the refracted light rays. By using the property that the axicon, when the point light source is on the optical axis, can be regarded as a double prism, he concluded that the caustic has two sheets; one is a two-dimensional surface, and the other is a segment of a line. For the case in which the point light source is placed off the optical axis, he explained that the sheet of the caustic corresponding to the segment of a line unfolds into a two-dimensional surface such that its transversal section has four cusp singularities.

The aim of the present work is to introduce a lens whose faces are a conical surface and a spherical surface. We illuminate this lens by a plane wavefront and its associated refracted wavefronts, light rays and caustic are computed. We find that the caustic has two branches. The first is constituted by two segments of a line, one part of this caustic is real and the other one virtual. The second branch of the caustic is a two-dimensional surface with a singularity of the cusp ridge type. It is important to remark that the two branches of the caustic are disconnected. Because of this property we believe that using this optical element one could generate a scalar optical accelerating beam in the region where the caustic is a two-dimensional surface of revolution, and at the same time a scalar optical beam with similar properties to the Bessel beam of zero order in the region were the real caustic is a segment of a line along the optical axis. 

To this end, in section\,2 we review the refraction of a plane wavefront by the plano spherical and the axicon lenses. That is, we assumed that we have a plane wavefront evolving along the $\hat{z}$ direction and it is refracted by a plano spherical lens. Then, we compute the corresponding refracted wavefronts, light rays and caustic. Furthermore, in this section we also obtain analogous results for the axicon lens. In section\,3 we study the refraction of a plane wave evolving along the $z$-axis by the conospherical lens. To be more precise, we obtain the refracted wavefronts, light rays and the corresponding caustic. We find that the caustic has three components, one of them is virtual, and the other two are real. The virtual caustic is a segment of a line along the $z$-axis and the two real branches of the caustic are a segment of a line along the optical axis and a two-dimensional surface of revolution about the $z$-axis, which has a singularity of the cusp ridge type. Finally, we present the conclusions.

\section{The plano spherical and the axicon lenses}
In this section we review the computation of the wavefronts, light rays and caustics associated with the refraction of a plane wavefront by a plano spherical and axicon lenses. These results allow us to introduce, in the next section, a conospherical lens. If we have a region of the free space filled out with two optical media with constant refractive indices, $n_1$ and $n_2$, separated by an interface given locally by a position vector $\mathbf{r}(\rho,\phi)$, then a plane wavefront evolving in the direction $\hat{z}$, in the optical medium with refractive index $n_1$, is refracted by the interface in the direction $\hat{\mathbf{R}}$. Thus, using the refraction law and the reference plane wavefront $\mathbf{s}=\rho(\cos\phi\hat{x} + \sin\phi\hat{y})$, a direct computation shows that the refracted wavefronts and light rays, are given by 
\begin{eqnarray}
\mathbf{X} & = & \mathbf{r}+ l\hat{\mathbf{R}}, \label{Wavefronts general} \\
l & = & \frac{\mathcal{C}}{n_{2}}-\gamma|\mathbf{r}-\mathbf{s}|, \label{l}  \\
\hat{\mathbf{R}} & = & \gamma \hat{z}+\Omega \hat{\mathbf{N}}, \label{R general} \\
\Omega & = & -\gamma(\hat{z} \cdot \hat{\mathbf{N}})+\sqrt{1-\gamma^{2}\left[1-(\hat{z} \cdot \hat{\mathbf{N}})^{2}\right]}, \label{Omega general}
\end{eqnarray}   
where $\mathcal{C}$ is a real parameter, $\hat{\mathbf{N}}$  is the unit vector to the interface and $\gamma \equiv n_1 / n_2$. It is important to remark that for fixed values of $\rho$ and $\phi$ equation (\ref{Wavefronts general}) describes a particular refracted light ray, while for a fixed value of $\mathcal{C}$ equations (\ref{Wavefronts general})-(\ref{R general}) describe a particular refracted wavefront.

From a mathematical point of view  equation (\ref{Wavefronts general}) describes a map between two subsets of ${\cal R}^3$, where ($\rho$, $\phi$, $l$) are local coordinates of the domain space and ($X$, $Y$, $Z$), with $\mathbf{X}=X\hat{x}+Y\hat{y}+Z\hat{z}$, are local coordinates of the target space. By definition, the \textit{focal region} associated with the refracted light rays \textit{is the caustic set} determined by the map (\ref{Wavefronts general}). The caustic set of a general map is computed by using the following:    

\textbf{Definition:} Let \textit{h:} ${\cal M}$ $\rightarrow$ ${\cal N}$ be a differentiable map, with ${\cal M}$ and ${\cal N}$ differentiable manifolds. The set of points in ${\cal M}$ where \textit{h} is not locally one-to-one is referred to as its critical set, and the image of the critical set is referred to as caustic set of $h$\,\cite{Arnold1, Arnold2, Arnold3}.

In accordance with this definition the critical set of the map (\ref{Wavefronts general}) is computed from the following condition
\begin{equation}
J \equiv  \left(  \frac{\partial \mathbf{X}}{\partial \phi} \right) \cdot \left[ \frac{\partial \mathbf{X}}{\partial l} \times  \frac{\partial \mathbf{X}}{\partial \rho} \right ] = H_2 l^2 + H_1 l + H_0 = 0,
\end{equation}
where
\begin{eqnarray}
H_0 & = &\hat{\mathbf{R}} \cdot \left[\left(\frac{\partial  \mathbf{r}}{\partial \rho} \right) \times \left(\frac{\partial  \mathbf{r}}{\partial \phi} \right)  \right], \nonumber \\
H_1 & = &\hat{\mathbf{R}} \cdot \left[\left(\frac{\partial  \mathbf{r}}{\partial \rho} \right) \times \left(\frac{\partial \hat{\mathbf{R}}}{\partial \phi} \right) + \left(\frac{\partial  \hat{\mathbf{R}}}{\partial \rho} \right) \times \left(\frac{\partial  \mathbf{r}}{\partial \phi} \right)  \right], \nonumber \\
H_2 & = &\hat{\mathbf{R}} \cdot \left[\left(\frac{\partial  \hat{\mathbf{R}}}{\partial \rho} \right) \times \left(\frac{\partial  \hat{\mathbf{R}}}{\partial \phi} \right)  \right]. \label{H0H1H2}
\end{eqnarray}
If $H_2\neq 0$, like in the plano spherical lens, the critical set has two branches given by
\begin{eqnarray}
l_{\pm}   = \frac{- H_1 \pm\sqrt{H_1^2 - 4 H_2 H_0}}{2 H_2}. \label{lpm}
\end{eqnarray} 
And the caustic set, which is obtained substituting equation (\ref{lpm}) into equation (\ref{Wavefronts general}),  can be written in the following manner 
\begin{eqnarray}
\mathbf{X}_{\pm}(\rho, \phi)  =  \mathbf{r} + \left(\frac{- H_1 \pm \sqrt{H_1^2 - 4 H_2 H_0}}{2 H_2} \right) \hat{\mathbf{R}}. \label{Caustica general 1}
\end{eqnarray}
However, if $H_2=0$ and $H_1\neq 0$, like in the axicon lens, then the critical set has only one branch given by 
\begin{equation}
l_c=\frac{-H_0}{H_1}. \label{lc}
\end{equation}
And the caustic set, which is obtained substituting equation (\ref{lc}) into equation (\ref{Wavefronts general}), has only one branch given by  
\begin{eqnarray}
\mathbf{X}_{c}(\rho, \phi)  =  \mathbf{r} - \left(\frac{H_0}{H_1}\right) \hat{\mathbf{R}}. \label{Caustica general 2}
\end{eqnarray}

\subsection{The plano spherical lens}
Now we assume that we have a plano spherical lens with refractive index $n$ in the free space. Using the polar coordinates $\rho$ and $\phi$, for this particular case we have  
\begin{eqnarray}
\mathbf{r}& = & \rho \hat{\rho} + \sqrt{a^2-\rho^2} \hat{z}, \label{r esfera} \\
\hat{\rho}& = & \cos \phi \, \hat{x} + \sin \phi \, \hat{y}, \label{unit rho}
\end{eqnarray}
where $0 \leq \rho \leq a $ and $0 \leq \phi \leq 2\pi $, with $a$ being the radius of the plane face. Using (\ref{r esfera}) and (\ref{unit rho}) a direct computation shows that 
\begin{eqnarray}
\hat{\mathbf{N}} = \frac{1}{a}\left( \rho \hat{\rho} + \sqrt{a^2-\rho^2} \hat{z}\right). \label{N esfera}
\end{eqnarray}
The plane wavefronts evolving in the direction $\hat{z}$ arrive to the plane surface of the lens and are transmitted without any reflection, then they arrive to the spherical surface and are refracted. Therefore, using equations (\ref{Wavefronts general})-(\ref{Omega general}) and (\ref{r esfera})-(\ref{N esfera}), a direct computation shows that the refracted wavefronts and light rays can be written in the following manner 
\begin{multline}
\mathbf{X} =  \left[ \frac{a^2 + l(\sqrt{a^2-\gamma^2\rho^2} - \gamma \sqrt{a^2-\rho^2)}}{a^2} \right] \rho \hat{\rho} \\
 +  \left[\frac{a^2\sqrt{a^2 - \rho^2} + l(a\gamma +\sqrt{a^2 - \rho^2})(\sqrt{a^2-\gamma^2\rho^2} - \gamma \sqrt{a^2-\rho^2)}}{a^2}\right]\hat{z},\label{WFLente plano esferico}
\end{multline}
where 
\begin{equation}
l = \frac{\mathcal{C}}{n_2} - \gamma \sqrt{a^2-\rho^2}.
\end{equation}
It is important to remark that equation (\ref{WFLente plano esferico}) is correct for $a^2-\gamma^2\rho^2 \geq 0$. That is, $0 \leq \rho \leq \rho_m = a/\gamma$.

\begin{figure}[!htp]
\begin{center}
\includegraphics[scale=.7]{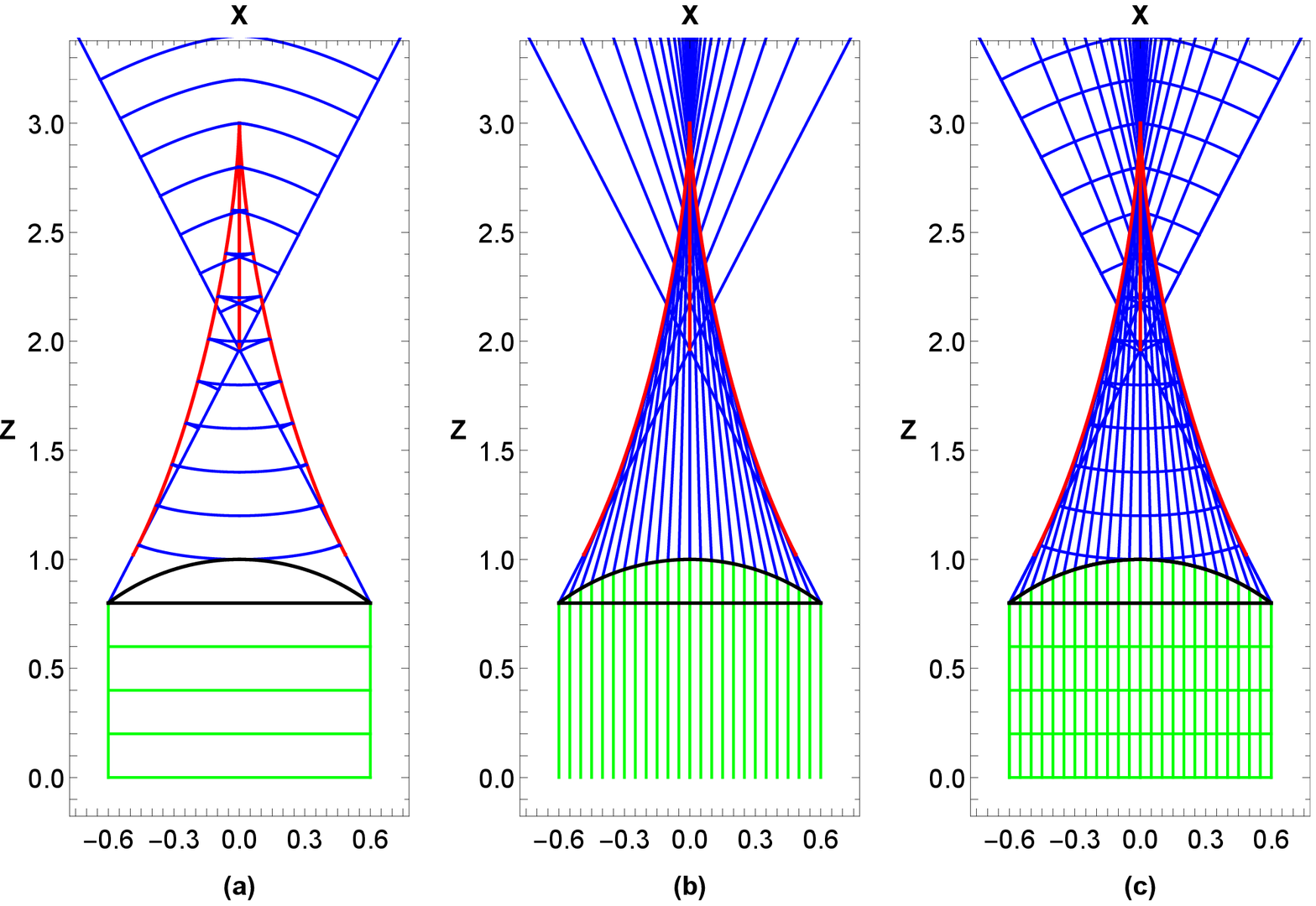}
\caption{\textit{\textit{Here we present on the plane $y=0$: (a) some incident plane wavefronts and the incident marginal light rays (Green), the plano spherical lens (Black), some refracted wavefronts and the refracted marginal light rays (Blue) and the two branches of the caustic (Red), (b) some incident light rays (Green), the plano spherical lens (Black), some refracted light rays (Blue) and the two branches of the caustic (Red), (c) superposition of the plots (a) and (b). To obtain these plots we use (\ref{WFLente plano esferico}), (\ref{Caustica menos esfera}) and (\ref{Caustica mas esfera}) with $a=1cm$ and $\gamma = 1.5$}}.}
\label{WF1}
\end{center}
\end{figure}

Finally, another direct computation shows that the two branches of the caustic region associated with the refracted light rays by the plane spherical lens can be written in the following manner:
\begin{eqnarray}
\mathbf{X}_{-} & = &  \frac{\gamma^2 \rho^3}{a^2}\hat{\rho} + \frac{\gamma  \left[(a^2-\gamma ^2 \rho ^2) ^{3/2} + \gamma (a^2-\rho ^2 )^{3/2}\right ] }{a^2 \left(\gamma ^2-1\right) } \hat{z}, \label{Caustica menos esfera} \\  
\mathbf{X}_{+} & = & \frac{\gamma  \left(\sqrt{a^2-\gamma ^2 \rho ^2}+\gamma  \sqrt{a^2-\rho ^2}\right)}{\gamma ^2-1} \hat{z}. \label{Caustica mas esfera}
\end{eqnarray}
Observe that $\mathbf{X}_{-}$ corresponds to a two-dimensional surface of revolution about the $ z $-axis with a degenerate singularity of the cusp type, while $\mathbf{X}_{+}$ corresponds to a segment of a line along the axis of symmetry of the optical system (see Figure\,\ref{WF1}).

\subsection{The plano conical lens}
\begin{figure}[!htp]
\begin{center}
\includegraphics[scale=.7]{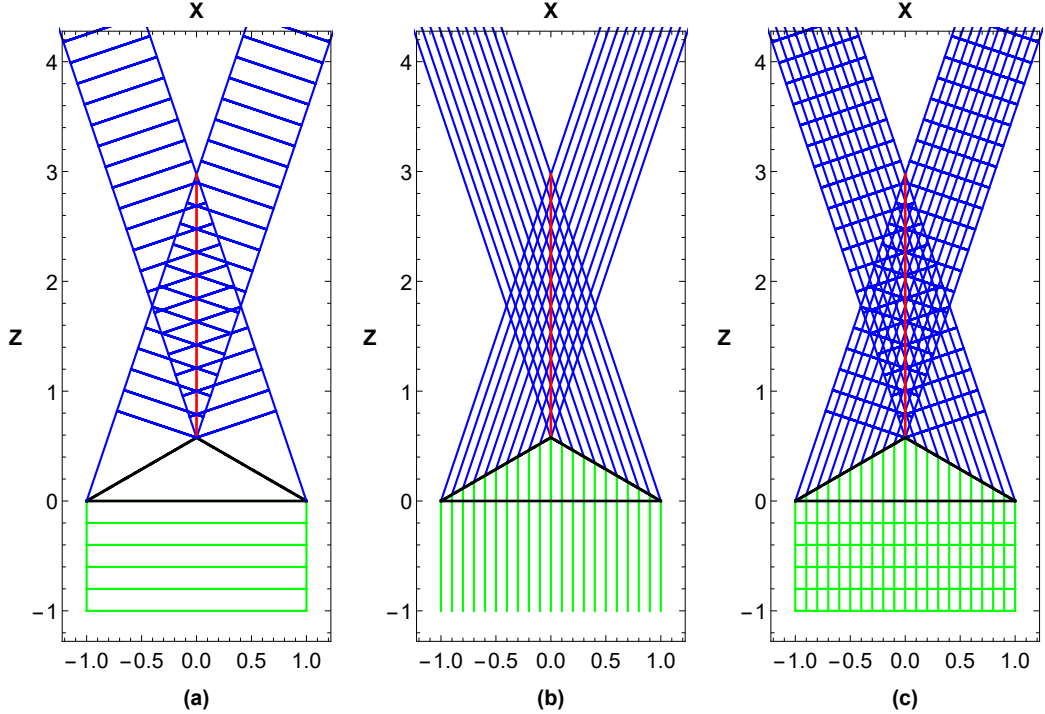}
\caption{\textit{\textit{Here we present on the plane $y=0$: (a) some incident plane wavefronts and the incident marginal light rays (Green), the plano conical lens (Black), some refracted wavefronts and the refracted marginal light rays (Blue) and the caustic (Red), (b) some incident light rays (Green), the plano conical lens (Black), some refracted light rays (Blue) and the caustic (Red), (c) superposition of the plots (a) and (b). To obtain these plots we use (\ref{WF lente planocono}) and (\ref{Caustica cono}) with $a=1cm$, $\alpha = \pi/3$ and $\gamma = 1.5$}}.}
\label{WF2}
\end{center}
\end{figure}
Now we assume that we have a plano conical lens with refractive index $n$ in the free space. Using the polar coordinates $\rho$ and $\phi$, for this particular case we have:
\begin{equation}
\mathbf{r} = \rho\hat{\rho} + (a-\rho)\cot \alpha\hat{z}, \label{r cono}
\end{equation}
where $0\leq \rho \leq a$, $0\leq \phi \leq 2\pi$ and $0\leq \alpha \leq \pi/2$, with $a$ being the radius of the plane face. Using (\ref{unit rho}) and (\ref{r cono}) a direct computation shows that 
\begin{equation}
\hat{\mathbf{N}} = \cos \alpha \hat{\rho} + \sin \alpha \hat{z}. \label{N cono}
\end{equation}
From equations (\ref{Wavefronts general})-(\ref{Omega general}), (\ref{unit rho}), (\ref{r cono}) and (\ref{N cono}), a direct computation shows that the refracted wavefronts and light rays when a plane wavefront is coming in the $\hat{z}$ direction can be written in the following manner
\begin{eqnarray}
\mathbf{X} & = & \left [ \rho + l \cos\alpha \left(\sqrt{1-\gamma^2\cos^2\alpha} - \gamma \sin \alpha \right ) \right] \hat{\rho} \nonumber \\&& + \left \{ (a-\rho) \cot \alpha +l \left[ \gamma + \sin \alpha \left(\sqrt{1-\gamma^2 \cos^2 \alpha}-\gamma \sin \alpha\right)\right] \right \} \hat{z},\label{WF lente planocono} 
\end{eqnarray}
where 
\begin{equation}
l = \frac{\mathcal{C}}{n_2} - \gamma (a-\rho) \cot \alpha.
\end{equation}
It is important to remark that equation (\ref{WF lente planocono}) is correct for $1-\gamma^2\cos \alpha \geq 0$. That is, $\arccos (1/\gamma) \leq \alpha < \pi/2$.

Finally, another direct computation shows that the caustic, associated with the refracted light rays by the plane conical lens, has only one branch; which can be written in the following manner:
\begin{equation}
\mathbf{X}_c = \left[\cot \alpha  (a-\rho  (\sin \phi +\cos \phi)) - \frac{\rho}{\Omega} \sec \alpha \left(\gamma + \Omega \sin \alpha\right) \right]\hat{z}, \label{Caustica cono}
\end{equation}
where 
\begin{equation}
\Omega = \sqrt{1-\gamma^2\cos^2 \alpha} - \gamma \sin \alpha. 
\end{equation}
Observe that $\mathbf{X}_c$ corresponds to a segment of line along the axis of symmetry of the optical system (see Figure \ref{WF2}).

\section{The conospherical lens}

\begin{figure}[!htp]
\begin{center}
\includegraphics[scale=2]{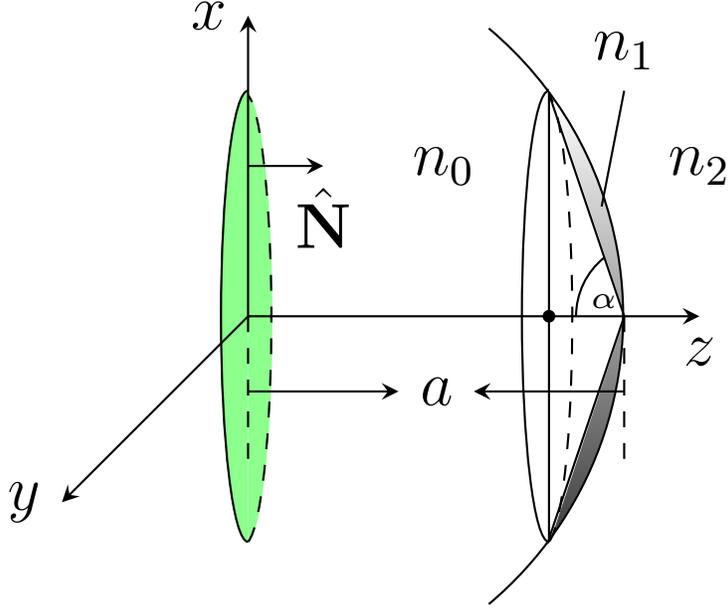}
\caption{\textit{\textit{Here we depict an incident plane wavefront and the conospherical lens with its corresponding parameters. The first surface is a cone given by (\ref{r1}) and the second surface is spherical and is given by (\ref{r2}), $a$ is the radius of the spherical surface and $\alpha$ is the aperture angle of the conical surface. The black region corresponds to the conospherical lens}}.}
\label{ConicalSphericalLens}
\end{center}
\end{figure}

In this section we compute the refraction of a plane wavefront by a conospherical lens, see Figure \ref{ConicalSphericalLens}. That is, we assume that a certain region of the free space is filled out with three optical media of refractive indices $n_0$, $n_1$ and $n_2$ respectively. The interface between the first and second media is a conical surface given by 
\begin{equation}
\mathbf{r}_1  =  \rho \hat{\rho} + (a-\rho\cot \alpha)\hat{z}, \label{r1}
\end{equation}
while the interface between the second and third media is a spherical surface given by 
\begin{eqnarray}
\mathbf{r}_2 & = & \rho_s \hat{\rho}_s + \sqrt{a^2-\rho_s^2} \hat{z},\label{r2}\\ 
\hat{\rho_s} & = & \cos \phi_s \hat{x} + \sin \phi_s \hat{y},
\label{ros}
\end{eqnarray}
where $\rho_s = \rho_s(\rho)$ and $\phi_s = \phi_s(\rho)$, are determined in the following manner. First, following the procedure used in subsection 2.2, we compute the refraction of the incoming plane wavefront by the conical surface of the lens given by $\mathbf{r}_1$. In this case a direct computation shows that the refracted light rays and wavefronts by this first surface of the lens can be written in the following manner 
\begin{eqnarray}
 \mathbf{X}_1  & = & \left[ \rho + l_1\cos \alpha \left(\sqrt{1-\gamma_0 ^2 \cos ^2 \alpha }-\gamma_0  \sin \alpha \right)  \right] \hat{\rho}\nonumber \\&& +  \left \{ a - \rho \cot \alpha + l_1 \left[ \gamma_0 + \sin \alpha \left( \sqrt{1-\gamma_0^2\cos^2 \alpha} - \gamma_0 \sin \alpha \right) \right ] \right \} \hat{z}, \label{X1}
\end{eqnarray}
where
\begin{equation}
l_1 = \frac{C_1}{n_1} - \gamma_0 \left( a - \rho \cot \alpha\right). 
\end{equation}
It is important to remark that for fixed values of $\rho$ and $\phi$, the parameter $C_1$ labels the points on a particular refracted light ray, while for fixed values of the parameter $C_1$, $\mathbf{X}_1$ describes a particular refracted wavefront. Observe that the points on the refracted light rays are also labeled by the parameter $l_1$. Now, we take the refracted light rays that arrive at the second spherical interface given by $\mathbf{r}_2$. To determine the relationship between the two sets of polar coordinates ($\rho$, $\phi$) and ($\rho_s$, $\phi_s$), from equations (\ref{r2}) and (\ref{X1}) we obtain the parametrization of the spherical surface of the lens $\mathbf{r}_2$ in terms of the polar coordinates ($\rho$, $\phi$) that parametrize the points on the first conical surface of the lens $\mathbf{r}_1$. To be more specific, we follow the refracted light rays by the first surface of the lens then when they arrive to the second surface of the lens we obtain the following relationship $\mathbf{X}_1=\mathbf{r}_2$, which determines the relationship we are looking for. This condition is equivalent to 
\begin{eqnarray}
\left[ \rho + l_1\cos \alpha \left(\sqrt{1-\gamma_0 ^2 \cos ^2 \alpha }-\gamma_0  \sin \alpha \right)  \right] \cos \phi  =  \rho_s \cos \phi_s, \label{X1=r21} \\
\left[ \rho + l_1\cos \alpha \left(\sqrt{1-\gamma_0 ^2 \cos ^2 \alpha }-\gamma_0  \sin \alpha \right)  \right] \sin \phi  =  \rho_s \sin \phi_s, \label{Y1=r22} \\
a - \rho \cot \alpha + l_1 \left[ \gamma_0 + \sin \alpha \left( \sqrt{1-\gamma_0^2\cos^2 \alpha} - \gamma_0 \sin \alpha \right) \right ]  =  \sqrt{a^2 - \rho_s^2}. \label{Z1=r23} 
\end{eqnarray} 
That is,
\begin{eqnarray}
\phi_s &= &\phi, \label{ro=ros} \\
l_1 & = & \frac{\sqrt{a^2 - \rho_s^2} + \rho \cot \alpha - a}{ \gamma_0 + \sin \alpha \left( \sqrt{1-\gamma_0^2\cos^2 \alpha} - \gamma_0 \sin \alpha \right)}, \label{l1roros}
\end{eqnarray} 
and 
\begin{eqnarray}
A\rho_s^2 + B\rho_s + C = 0,
\end{eqnarray}
where
\begin{eqnarray}
A & = & -\frac{\sec^2 \alpha (  2\gamma_0 
   \Omega_1 \sin \alpha + \gamma_0^2+ \Omega_1^2 )}{\Omega_1^2}, \nonumber \\
B & = & \frac{2 (\Omega_1 + \gamma_0 \csc \alpha) [\rho(\gamma_0 + \Omega_1 \csc \alpha) - a \Omega_1 \cos \alpha ] \sec \alpha \tan \alpha}{\Omega_1^2}, \nonumber \\
C & = & -\frac{\rho (\gamma_0 + \Omega_1 \csc\alpha)(\rho(\gamma_0 + \Omega_1 \csc \alpha )-2 a \Omega_1 \cos \alpha)\sec^2 \alpha}{\Omega_1^2}, \nonumber \\
\Omega_1 & = & \sqrt{1-\gamma_0^2\cos^2 \alpha} - \gamma_0 \sin \alpha, 
\end{eqnarray}
with $\gamma_0 \equiv n_0/n_1$. Therefore, the two solutions for $\rho_s$ are explicitly given by  
\begin{equation}
\hspace{-2cm} \rho_s = \rho_{\pm} =  \frac{\sin \alpha  \left \{ (\Omega_1 + \gamma_0 \csc \alpha)[ a \Omega_1 \cos \alpha -\rho (\gamma_0 + \Omega_1 \csc \alpha)]    \pm \csc \alpha \sqrt{\Delta} \right \}}{\gamma_0^2 + 2\gamma_0 \Omega_1 \sin \alpha + \Omega_1^2}, \label{rosmm}
\end{equation}
with 
\begin{eqnarray}
\Delta &  = & (\Omega_1 + \gamma_0 \csc \alpha)^2 [a \Omega_1 \cos \alpha   - \rho(\gamma_0 + \Omega_1 \csc \alpha)]^2 \sin^2 \alpha  - \rho (\gamma_0 + \Omega_1 \csc \alpha) \nonumber \\
& & \times [ \rho (\gamma_0 + \Omega_1 \csc \alpha)  - 2 a \Omega_1 \cos \alpha]
 ( \gamma_0^2 + 2 \gamma_0 \Omega_1 \sin \alpha + \Omega_1^2 ).
\end{eqnarray}

Until this moment we have computed the refraction of the incoming plane wavefront evolving along the $z$-axis by the first conical surface of the lens and we have obtained the parameterization of the second spherical surface of the lens in terms of the polar coordinates ($\rho$, $\phi$), which label the points on the first conical surface of the lens. Thus, the parametrization of the second spherical surface that we have to use to compute the second refraction of the light rays is given  by  
\begin{equation}
\mathbf{r}_2 = \rho_s \hat{\rho} + \sqrt{a^2-\rho_s^2} \hat{z},\label{r2s}
\end{equation}
where $\rho_s$ is given by (\ref{rosmm}). 

Therefore, the refracted light rays and wavefronts by the second surface of the lens (\ref{r2s}) can be written in the following manner 
\begin{eqnarray}
\mathbf{X}_{2} & = & \mathbf{r}_{2}+\left[\frac{\mathcal{C}_{2}}{n_{2}}-\gamma_{0} \gamma_{1}\left|\mathbf{r}_{1}-\mathbf{s}\right|-\gamma_{1}\left|\mathbf{r}_{2}-\mathbf{r}_{1}\right|\right] \hat{\mathbf{R}}_{2},\label{X2}  \\
\hat{\mathbf{R}}_2 & = & \gamma_1 \hat{\mathbf{I}}_2 +\Omega_2 \hat{\mathbf{N}}_2, \label{R2}\\
\Omega_2 & = & -\gamma_1(\hat{\mathbf{I}}_2 \cdot \hat{\mathbf{N}}_2)+\sqrt{1-\gamma_1^{2}\left[1-(\hat{\mathbf{I}}_2 \cdot \hat{\mathbf{N}}_2)^{2}\right]}, \label{Omega2} 
\end{eqnarray}
where  
\begin{eqnarray}
\hat{\mathbf{I}}_2 & = &\Omega_1 \cos \alpha \hat{\rho} + \left( \gamma + \Omega_1 \sin \alpha \right )\hat{z}, \\
\hat{\mathbf{N}}_2 & = & \frac{\rho_s}{a}\hat{\rho}_s + \frac{\sqrt{a^2-\rho_s^2}}{a} \hat{z}. \label{N2}
\end{eqnarray}
Using equations (\ref{R2})-(\ref{N2}) a direct computation shows that 
\begin{eqnarray}
\hat{\mathbf{R}}_2 & = & \left( \gamma_1 \Omega_1 \cos \alpha + \frac{\Omega_2 \rho_s}{a} \right) \hat{\rho} + \left(\gamma_1 (\gamma_0 + \Omega \sin \alpha ) + \frac{\Omega_2 \sqrt{a^2-\rho_s^2}}{a} \right)\hat{z}, \label{R2unit} \\
\Omega_2 & = & - \frac{\gamma_1 \rho_s \Omega_1 \cos \alpha}{a} - \frac{\gamma_1 \sqrt{a^2-\rho_s^2} (\gamma_0 + \Omega_1 \sin \alpha )}{a} + \frac{\sqrt{\Delta_2}}{a}, \\
\Delta_2 & = & a^2- \gamma_1 \left(a^2 - \left(\rho_s \Omega_1 \cos \alpha + \sqrt{a^2-\rho_s^2}(\gamma_0 + \Omega_1\sin \alpha \right)^2 \right).
\end{eqnarray}
Since $\mathbf{s}=\rho \hat{\rho}$, finally using equations (\ref{r1}), (\ref{r2s}) and (\ref{R2unit}) another  direct computation shows that equation (\ref{X2}) can be written in the following manner 
\begin{eqnarray}
\mathbf{X}_2 & = & \left[ \rho_s + l_2 \left( \gamma_1 \Omega_1 \cos \alpha + \frac{\Omega_2 \rho_s}{a} \right) \right ]\hat{\rho} \nonumber \\
& & +\left[ \sqrt{a^2 - \rho_s^2} \left( 1 + \frac{l_2\Omega_2}{a} \right) + l_2 \gamma_1 \left( \gamma_0 + \Omega_1 \sin \alpha   \right) \right ]\hat{z},\label{X2F}
\end{eqnarray}
where
\begin{equation}
\hspace{-2cm} l_{2}= \frac{\mathcal{C}_{2}}{n_{2}} - \gamma_0 \gamma_1 (a-\rho \cot \alpha) - \gamma_1 \sqrt{(\rho-\rho_s)^2 + \left( \sqrt{a^2-\rho_s^2} - a + \rho \cot \alpha\right)^2}.
\end{equation}

In conclusion, equation (\ref{X2F}) provides both the light rays and wavefronts refracted by the second spherical surface of the lens. Since (\ref{X2F}) describes a map between two subsets of ${\cal R}^3$, where ($\rho$, $\phi$, $l_2$) are local coordinates in the domain space and ($X_2$, $Y_2$, $Z_2$), with $\mathbf{X}_2 = X_2\hat{x} +  Y_2 \hat{y} + Z_2 \hat{z}$, are local coordinates in the target space. Therefore, the critical set associated with this map is computed from the following condition 
\begin{equation}
J \equiv  \left(  \frac{\partial \mathbf{X}_2}{\partial \phi} \right) \cdot \left[ \frac{\partial \mathbf{X}_2}{\partial l_2} \times  \frac{\partial \mathbf{X}_2}{\partial \rho} \right ] = 0.
\end{equation}
Since 
\begin{eqnarray}
\frac{\partial \mathbf{X}_2}{\partial \phi} & = & D \hat{\phi}, \\
\frac{\partial \mathbf{X}_2}{\partial l_2} & = & E\hat{\rho} + F\hat{z}, \\
\frac{\partial \mathbf{X}_2}{\partial \rho} & = & G\hat{\rho} + H\hat{z},
\end{eqnarray}
where
\begin{eqnarray}
D & = &  \rho_s + l_2 \left( \gamma_1 \Omega_1 \cos \alpha + \frac{\Omega_2 \rho_s}{a} \right),  \\
E & = &   \gamma_1 \Omega_1 \cos \alpha + \frac{\Omega_2 \rho_s}{a},  \\
F & = &  \frac{\Omega_2 \sqrt{a^2 - \rho_s^2}}{a}  +  \gamma_1 \left( \gamma_0 + \Omega_1 \sin \alpha   \right), \\
G & = &  \rho_s' + l_2  \frac{(\Omega_2 \rho_s)'}{a} , \\
H & = &  l_2 \left( \frac{\Omega_2 \sqrt{a^2 - \rho_s^2}}{a} \right)', 
\end{eqnarray}
and the prime denotes the derivative with respect to the coordinate $\rho$. Therefore, the critical set of the map (\ref{X2F}) is determined from the following condition 
\begin{equation}
J = D(FG-EH)= 0,
\end{equation}
which implies that the two branches of the critical set are given by $D=0$ and $FG=EH$. That is 
\begin{eqnarray}
l_{2} & = &  l_{21} = \frac{-a\rho_s}{a \gamma_1 \Omega_1 \cos \alpha + \Omega_2 \rho_s },\label{l21} \\
l_{2} & = &  l_{22}=\frac{aF\rho_s'}{E(\Omega_2 \sqrt{a^2-\rho_s^2})'-F(\Omega_2 \rho_s)'}. \label{l22}
\end{eqnarray}
To determine the two branches of the caustic region we substitute (\ref{l21}) and (\ref{l22}) into (\ref{X2F}). Thus, we obtain 
\begin{eqnarray}
\mathbf{X}_{c21} & = & \left[ \sqrt{a^2 - \rho_s^2} \left( 1 + \frac{l_{21}\Omega_2}{a} \right) + l_{21} \gamma_1 \left( \gamma_0 + \Omega_1 \sin \alpha   \right) \right ]\hat{z},\label{Xc21}\\
\mathbf{X}_{c22} & = & \left[ \rho_s + l_{22} \left( \gamma_1 \Omega_1 \cos \alpha + \frac{\Omega_2 \rho_s}{a} \right) \right ]\hat{\rho} \nonumber \\
& & +\left[ \sqrt{a^2 - \rho_s^2} \left( 1 + \frac{l_{22}\Omega_2}{a} \right) + l_{22} \gamma_1 \left( \gamma_0 + \Omega_1 \sin \alpha   \right) \right ]\hat{z}.\label{Xc22}
\end{eqnarray}
Observe that $\mathbf{X}_{c21}$ describes a segment of a line along the $z$-axis and $\mathbf{X}_{c22}$ describes a revolution surface about the $z$-axis. In Figure\,\ref{WF3} we find that $\mathbf{X}_{c21}$ corresponds to two disconnected segments of line, one is real and the other one is virtual. 

\section{A particular example of a conospherical lens}
\begin{figure}[!htp]
\begin{center}
\includegraphics[scale=1]{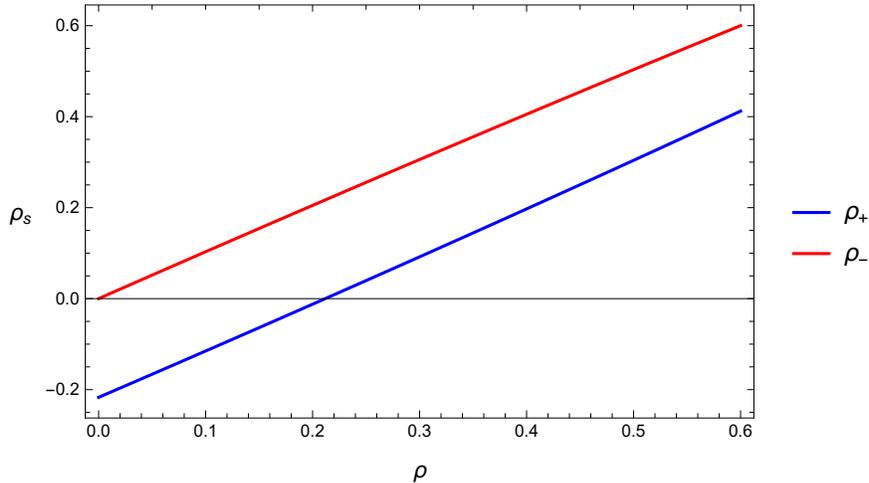}
\caption{\textit{\textit{Here we plot $\rho_s=\rho_{\pm}$ given by (\ref{rosmm})}.}}
\label{ros+-}
\end{center}
\end{figure}
\begin{figure}[!htp]
\begin{center}
\includegraphics[scale=1.3]{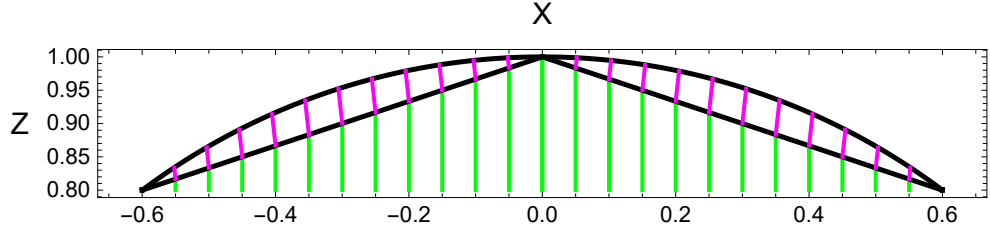}
\caption{\textit{\textit{In this plot we present some incident light rays (green), the conical surface (black), the corresponding refracted light rays inside the lens (magenta) and the spherical surface (black)}}.}
\label{RayosInCIn}
\end{center}
\end{figure}
\begin{figure}[!htp]
\begin{center}
\includegraphics[scale=.6]{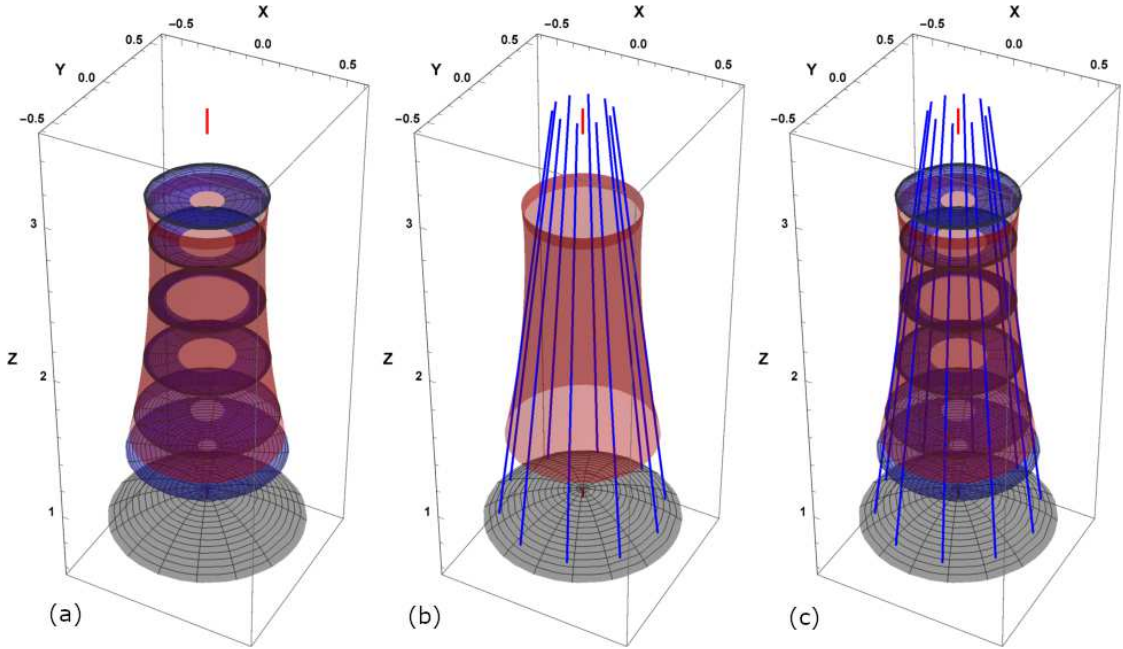}
\caption{\textit{\textit{In this plot we present: (a) the conospherical lens (gray), some refracted wavefronts leaving the lens (blue) and the branches of the caustic (red). (b) the conospherical lens (gray), some refracted light rays leaving the lens (blue) and the branches of the caustic (red). Finally, in (c) we present the superposition of the plots (a) and (b)}}.}
\label{3DConicalSphericalLens}
\end{center}
\end{figure}

In this section we apply the general results to a particular conospherical lens characterized by $a=1$, $\gamma_0=2/3$, $\alpha= {\rm arccot}\,(1/3)$, $ \gamma=3/2$. That is, we are assuming that we have a conospherical lens with refractive index $n_1=3/2$ in free space ($n_0=1$ and $n_2=1$). Furthermore, from (\ref{WFLente plano esferico}) we find that in this case $0\leq \rho \leq \rho_m =a/\gamma_0 = 2/3$. Using these particular values for the parameters and equation (\ref{rosmm}) in Figure\,\ref{ros+-} we present the plots of $\rho_s$. From this plot we find that the right expression for $\rho_s$ is given by $\rho = \rho_-$. Therefore, the refracted light rays inside the lens are given by 
\begin{figure}[!htp]
\begin{center}
\includegraphics[width = 15.5cm]{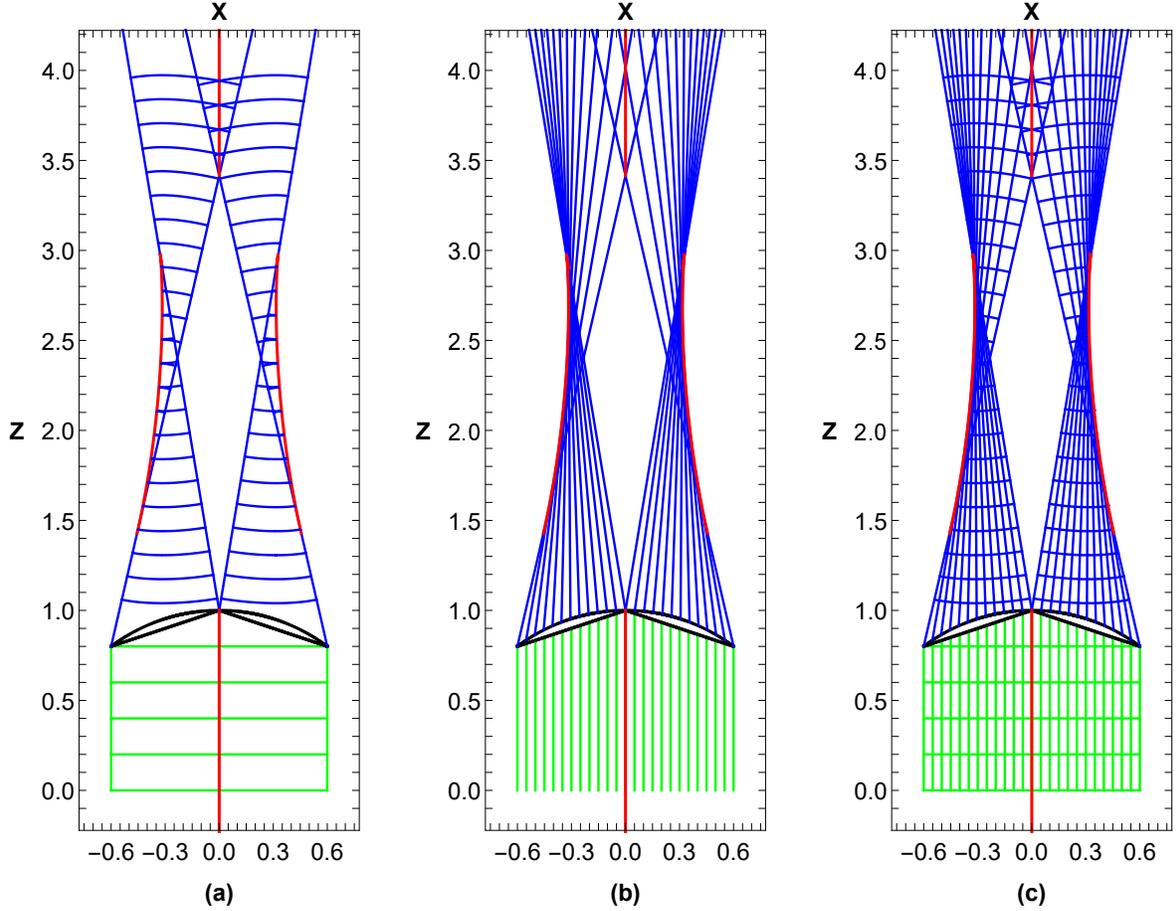}
\caption{\textit{\textit{Here we present on the plane $y=0$: (a) some incident plane wavefronts and the incident marginal light rays (Green), the plano spherical lens (Black), some refracted wavefronts and the refracted marginal light rays (Blue) and the two branches of the caustic (Red), (b) some incident light rays (Green), the plano spherical lens (Black), some refracted light rays (Blue) and the two branches of the caustic (Red), (c) superposition of the plots (a) and (b). To obtain these plots we use equations (\ref{r1}), (\ref{r2s}), (\ref{X2F}), (\ref{Xc21}) and (\ref{Xc22})}}.}
\label{WF3}
\end{center}
\end{figure}
\begin{equation}
\textbf{X}_i = \mathbf{r}_1 + \sigma(\mathbf{X}_{1-}- \mathbf{r}_1), \label{Xi}
\end{equation}
where $0 \leq \sigma \leq 1$, 
\begin{eqnarray}
\mathbf{X}_{1-}  & = & \left[ \rho + \cos \alpha \left(\sqrt{1-\gamma_0 ^2 \cos ^2 \alpha }-\gamma_0  \sin \alpha \right)  \right] \hat{\rho}\nonumber \\&&  \hspace{-0.5cm}+  \left \{ a - \rho \cot \alpha + l_{\rho} \left[ \gamma_0 + \sin \alpha \left( \sqrt{1-\gamma_0^2\cos^2 \alpha} - \gamma_0 \sin \alpha \right) \right ] \right \} \hat{z}, \label{X1-}
\end{eqnarray}
and 
\begin{eqnarray}
l_{\rho} & = & \frac{\sqrt{a^2 - \rho_{-}^2} + \rho \cot \alpha - a}{ \gamma_0 + \sin \alpha \left( \sqrt{1-\gamma_0^2\cos^2 \alpha} - \gamma_0 \sin \alpha \right)}. \label{l1roros-}
\end{eqnarray} 
Thus, using equations (\ref{Xi})-(\ref{l1roros-}) in Figure\,\ref{RayosInCIn} we present some incident light rays, the conospherical lens and the corresponding refracted light rays inside the lens. In Figure\,\ref{WF3} we present on the plane $y=0$: (a) some incident wavefronts and the incident marginal light rays, the conospherical lens, the corresponding refracted wavefronts leaving the lens and the branches of the caustic. In (b) we present some incident light rays, the conospherical lens, the corresponding refracted light rays leaving the lens and the branches of the caustic. Finally, in (c) we show the superposition of the plots (a) and (b). From these plots it is clear that the caustic, for this case, has three branches two of them are real and one virtual. The virtual branch of the caustic is behind the lens and the real ones are in front the lens. One of the real branches corresponds to a segment of a line along the $z$-axis, which is similar to the caustic generated by the axicon lens. The second real branch of the caustic is a two-dimensional surface of revolution about the $z$-axis with a singularity of the cusp ridge type. Finally, using equations (\ref{r1}), (\ref{r2s}), (\ref{X2F}), (\ref{Xc21}) and (\ref{Xc22}) in Figure\,\ref{3DConicalSphericalLens} we present (a) the conospherical lens, some refracted wavefronts leaving the lens and the branches of the caustic, in (b) we present the conospherical lens, some refracted light rays leaving the lens and the branches of the caustic. Finally, in (c) we show the superposition of the plots (a) and (b). In these plots we show the two real branches of the caustic, which correspond to a segment of a line along the $z$-axis and a two dimensional surface of revolution about the $z$-axis with a singularity of the cusp ridge type.         

\section{Conclusions}
In this work we have introduced the conospherical lens; that is, a lens whose faces are a conical surface and a spherical surface. Then we illuminated this lens with a plane wavefront and its associated refracted wavefronts, light rays and caustic were computed. We found that the caustic has two branches. The first is constituted by two segments of a line, one part of this caustic is real and the other one virtual. The second branch of the caustic is a two-dimensional surface with a singularity of the cusp ridge type. It is important to remark that the two branches of the caustic are disconnected, see Figure\,\ref{WF3}, and because of this property we believe that using this optical element one could generate a scalar optical accelerating beam in the region where the caustic is a two-dimensional surface of revolution, and at the same time a scalar optical beam with similar properties to the Bessel beam of zero order in the region were the real caustic is a segment of a line along the optical axis. That is, this optical element could be used to generate a scalar optical beam such that in a region is accelerating and in the other behaves like the Bessel beam of zero order. These properties are under study and the results will be reported elsewhere.

\begin{acknowledgments}
Gilberto Silva-Ortigoza acknowledges financial support from SNI. 
\end{acknowledgments}


\end{document}